\documentclass[prl,twocolumn,lengthcheck,superscriptaddress]{revtex4}

\usepackage{amsmath}
\usepackage{times}
\usepackage{amssymb}
\usepackage{graphicx}
\usepackage{bm}

\newcommand{\bra}[1]{\left\langle{#1}\right\vert}
\newcommand{\ket}[1]{\left\vert{#1}\right\rangle}

\newcommand{\aop}[1]{a^{\phantom\dag}_{#1}}
\newcommand{\adagop}[1]{a^\dag_{#1}}
\newcommand{\ee}{\mathbb{E}}
\newcommand{\id}{\mathbb{I}}

\begin{document}

\title{Information propagation through quantum chains with fluctuating disorder}

\author{Christian K.\ \surname{Burrell}}
\affiliation{Department of Mathematics, Royal Holloway University of London, Egham, Surrey, TW20 0EX, UK}

\author{Jens \surname{Eisert}}
\affiliation{Institute of Physics and Astronomy, University of Potsdam, 14476 Potsdam,
Germany}
\affiliation{QOLS, Blackett Laboratory, Imperial College London, Prince Consort Road, London SW7 2BW, UK} 

\author{Tobias J.\ \surname{Osborne}}
\affiliation{Department of Mathematics, Royal Holloway University of
London, Egham, Surrey, TW20 0EX, UK}

\begin{abstract}
We investigate the propagation of information through one-dimensional quantum chains in fluctuating external fields. We find that information propagation is suppressed, but in a quite different way compared to the situation with static disorder. We study two settings: (i) a general model where an unobservable fluctuating field acts as a source of decoherence; (ii) the XX model with both observable and unobservable fluctuating fields. In the first setting we establish a noise threshold below which information can propagate ballistically and above which information is localised. In the second setting we find localisation for all levels of unobservable noise, whilst an observable field can yield diffusive propagation of information.
\end{abstract}

\maketitle

Quantum lattice models as frequently encountered in the condensed-matter context or in quantum optics obey a kind of locality: once locally excited the excitation will travel through the lattice at a finite velocity. For spin models this speed at which information can propagate is generally limited by the \emph{Lieb-Robinson bound} \cite{LieRob72} which says that there is an effective \emph{``light cone''} (or \emph{``sound cone''}) for correlations, with exponentially decaying tails, whose radius grows \emph{linearly} with time \cite{NacSim05}. The importance of understanding information propagation in interacting quantum systems was only understood recently when it was exploited to establish the Lieb-Schultz-Mattis theorem in higher dimensions \cite{Has04}. In generalising the proof of this breakthrough result an intimate link between the speed of
information propagation and the efficient simulation of these systems has been revealed \cite{Osb05,Has08}.

The argument underlying the Lieb-Robinson bound relies only on the
ultra-violet cutoff imposed by the lattice structure and is
therefore very general. Hence there are some situations where the
Lieb-Robinson bound is not the best available: when we know more
about the structure of the interactions it should be possible to
construct tighter bounds---an intuition which has been borne out
in Refs.\ \cite{BurOsb07, ZniProPre07} where it was shown that for the
XX model with static disorder there exists an effective
``light cone'' whose radius grows \emph{logarithmically} with time
\cite{BurOsb07}. In this example \emph{Anderson localisation}
\cite{And57} for the tight-binding model was combined with a
Lieb-Robinson type argument to supply the stronger bound on the
multiparticle dynamics.

 \begin{figure}
  \centering
  \includegraphics[width=\columnwidth]{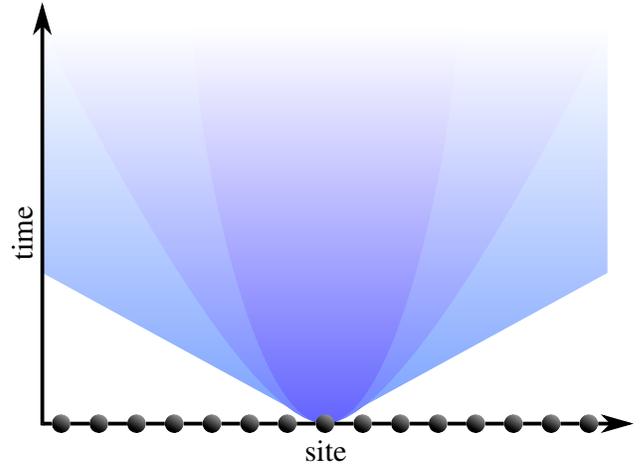}
  \caption{Schematic illustration of the different ``light cones'' for 
   the different regimes: (a) the (lower) \emph{linear} one
   for ordered systems; (b) the (upper) \emph{logarithmic} one 
   for static disorder; (c) the (middle) \emph{diffusive} one for dynamic disorder. In each regime information is strongly attenuated outside the associated light cone.} \label{fig:light-cones}
 \end{figure}

Interacting quantum lattice systems are also of great interest in quantum information science as they could be used as \emph{quantum wires} to join together different parts of quantum computers or quantum devices \cite{Bur06, Bos08}. When used in this manner quantum spin chains become \emph{quantum channels} and the standard Lieb-Robinson bound then supplies a fundamental lower bound on the time required to transmit information: the receiver must wait for a time at least proportional to the length of the chain before the message arrives. For the disordered XX model mentioned above the bound of Ref.\ \cite{BurOsb07} shows that the receiver must wait for a time which is \emph{exponential} in the chain length before information about the message can be extracted, thus rendering such systems essentially useless as quantum channels.

In this work we obtain bounds on the propagation of information through
interacting quantum spin chains in a \emph{fluctuating} disordered field, a setting which is ubiquitous in the condensed matter context \cite{CM}. It will become apparent that---in some situations---information propagates diffusively; in other situations the system becomes localised in the sense that information can propagate effectively by at most a constant number of sites.

We study two models: (i) a spin chain with general nearest-neighbour interactions in a noisy field fluctuating both in strength and direction; (ii) an XX spin chain in a field fluctuating in strength only. To solve the dynamics of both these models we find a master equation which describes the time evolution. We then explicitly calculate an improved Lieb-Robinson bound for the first model before finding various correlation functions for the second model. This allows us to identify several different possible regimes of information propagation: either localised or ballistic propagation for the first model and either localised or diffusive propagation for the second.

{\it Master equation ---} We derive a master equation for each of 
our models with time-dependent Hamiltonian  
 \begin{eqnarray}\label{eqn:hamiltonian}
  H(\xi,t) = H_0 + \sum_{\alpha,j} \xi^\alpha_j(t) \sigma^\alpha_j
 \end{eqnarray}
where $\sigma^x, \sigma^y, \sigma^z$ are the Pauli matrices, 
$H_0$ is the intrinsic system Hamiltonian and the second term describes the interaction with the noisy field. For the first model $\alpha$ runs over the three directions $x,y,z$ whilst for the second model the field direction is fixed ($\alpha=z$); $j$ runs over all lattice sites and $\xi_j^\alpha(t)$ are the field strengths at time $t$, which must be distributed according to a probability distribution with finite second moment (e.g. a Gaussian distribution, in which case $\xi_j^\alpha = dB_j^\alpha$ describes a Brownian motion $B_j^\alpha$).

After time $t$ an initial state $\rho(0)$ evolves to $\rho(\xi,t) = U(\xi,t)\rho(0)U(\xi,t)^\dag$ where the propagator $U(\xi,t)$ is
given by the time-ordered exponential $U(\xi,t)=\mathcal{T}\exp{(-i\int_0^t H(\xi,s)ds)}$. We now describe the derivation of the master equation for the \emph{ensemble averaged density operator} $\rho(t) = \mathbb{E}_\xi  \rho(\xi,t)$ where $\mathbb{E}_\xi$ represents an average over all possible realisations of the disorder. We split the time evolution up into $M$ small time steps and at each step we couple each site to ancillas  or meters---one for each direction $x,y,z$ (only one meter per site is used for the model with fixed field direction). When each meter is initialised in a Gaussian state, one can take the small time step limit $M\to\infty$, following the steps of continuous-measurement theory in \cite{CavMil87}, resulting in the  master equation
 \begin{equation}\label{eqn:master1}
  \partial_t \rho(t) = -i[\rho(t), H_0] - \gamma \sum_{\alpha,j}
  [[\rho(t),\sigma^\alpha_j],\sigma^\alpha_j]
 \end{equation}
where $\gamma = c\ee(\xi^\alpha_j)^2$, $c>0$, is a constant measuring the strength of the disorder. As before, for the second model we omit the sum over $\alpha$ and simply set $\alpha=z$. The above derivation is valid when all times scales of interest are longer than the typical fluctuation timescale.

{\it General spin chain model ---} In this section we derive a Lieb-Robinson bound for a general spin chain of length $n$ in an unobservable field whose strength and direction fluctuate independently on each site. The Hamiltonian is given by Eq.\ (\ref{eqn:hamiltonian}) with intrinsic Hamiltonian equal to a sum of general nearest-neighbor interactions $H_0 = \sum_j h_j$, where $h_j$ acts on sites $j$ and $j+1$. Note that this model is general enough to include all the standard quantum lattice systems, including the ferromagnetic and antiferromagnetic Heisenberg models and also frustrated systems.

A Lieb-Robinson bound governs the speed of ``light'' or ``sound'' in the lattice model. It is commonly expressed as an upper bound on the \emph{Lieb-Robinson commutator},
 \begin{equation*}
  C_B(x,t) = \sup_{A_x} \frac{\|[A_x,B(t)]\|}{\|A_x\|}
 \end{equation*}
where $A_x$ is an operator acting non-trivially only on site $x$, $B(t) =\ee_\xi  U^\dag(\xi,t) B U(\xi,t)$ is an ensemble average over all possible realisations of the operator $B$ in the Heisenberg picture, and $\|\cdot\|$ is the operator norm. While the Lieb-Robinson commutator is not directly observable, any bound for it easily translates to a bound on \emph{all} observable time-dependent two-point correlation functions $\langle A B(t) \rangle - \langle A\rangle \langle B(t)\rangle$. Hence, the Lieb-Robinson bound is a convenient bound to express the decay of correlations.

In the Heisenberg picture $B$ obeys an equation similar to the master equation derived in the previous section, namely $\partial_t B(t) = i[H_0,B(t)] - \gamma \sum_j \mathcal{D}_j(B(t))$ where the decoherence operators are defined as  $\mathcal{D}_j(B(t))=\sum_\alpha[\sigma^\alpha_j,[\sigma^\alpha_j,B(t)]]$. Defining $\mathcal{D}(B) = \sum_j \mathcal{D}_j(B)$ and using a Taylor expansion we arrive at
 \begin{equation*}
  \begin{array}{rcl}
   \|[A_x,B(t)]\| & \leq & (1-8\varepsilon\gamma n) \, \|[A_x,B(t)+i\varepsilon[H_0,B(t)]]\|\\
   & + & \varepsilon\gamma\|[A_x,8 n B(t)-\mathcal{D}(B(t))]\| + O(\varepsilon^2)
  \end{array}
 \end{equation*}
We deal with each of the non-negligible terms on the right hand side
in turn. Making repeated use of Taylor expansions and the unitary
equivalence of operator norm it is easy to see that the first term
can bounded as $\|[A_x,B(t)+i\varepsilon[H_0,B(t)]]\| \leq \varepsilon \|[[H_0,A_x],B(t)]\| + \|[A_x,B(t)]\| + O(\varepsilon^2)$. In order to bound the second term, we note the following two identities: (i) $\mathcal{D}_k(\sigma^\alpha_j) = 8\delta_{j,k}\sigma^\alpha_k$ 
for all $\alpha\neq 0$ and $\mathcal{D}_k(\sigma^0_j) = 0$, where $\sigma^0_k = \id$; (ii) for all $\beta\neq 0$
 \begin{equation}
  \sum_{\alpha\in\{0,x,y,z\}} \sigma^\alpha_k \sigma^\beta_k
  \sigma^\alpha_k = 0,\,
  \sum_{\alpha\in\{0,x,y,z\}} \sigma^\alpha_k \sigma^0_k
  \sigma^\alpha_k = 4\id
 \end{equation}
If we define $B_{\bm\alpha} = \sigma^{\alpha_1}_1 \otimes\cdots\otimes
\sigma^{\alpha_n}_n$ to be a tensor product operator indexed by the
vector $\bm\alpha = (\alpha_1, \ldots, \alpha_n)$, it is straightforward to use the above identities to show that
 \begin{equation}\label{eqn:commutator2}
  \begin{array}{l}
   \|[A_x,8nB_{\bm\alpha} - \mathcal{D}(B_{\bm\alpha})]\| \\
   \qquad \leq 
   \sum_{k\neq x,\beta\in\{0,x,y,z\}} \|[A_x,\sigma^\beta_k B_{\bm\alpha} \sigma^\beta_k]\|\\
   \qquad = 8(n-1)\|[A_x,B_{\bm\alpha}]\|
  \end{array}
 \end{equation}
We can extend this result to general $B$ by linearity: $B(t) = \sum_{\bm\alpha} c_{\bm\alpha}(t)B_{\bm\alpha}$. This implies that the second term obeys the bound (\ref{eqn:commutator2}) with $B_{\bm\alpha}$ replaced by $B(t)$. Combining the above equations allows us to bound the time derivative $\partial_t C_B(x,t) \leq -8\gamma C_B(x,t) + \sup_{A_x} {\|[[H_0,A_x],B(t)]\|}/{\|A_x\|}$. It
remains to rewrite the off-diagonal term, the right-most term containing the double commutator. Note that $[H_0,A_x] = 2\|H_0\|\|A_x\|V$ with $V = \sum_{\alpha,\beta\in\{0,x,y,z\}} (u^{\alpha,\beta}_{x-1}\sigma^\alpha_{x-1}\sigma^\beta_x + u^{\alpha,\beta}_x \sigma^\alpha_x\sigma^\beta_{x+1} )$ where $|u^{\alpha,\beta}_y| \leq \|V\| \leq 1$. By defining the vector $\textbf{\emph{C}}_B(t) = (C_B(1,t), \ldots, C_B(n,t))$ and employing the above expansion for $V$ we are able to rewrite the bound on the time-derivative of the Lieb-Robinson commutator as
 \begin{equation*}
  \partial_t \textbf{\emph{C}}_B(t) \leq (-8(\gamma-8\|H_0\|)\id +32\|H_0\|R)
  \textbf{\emph{C}}_B(t)
 \end{equation*}
where the inequality is to be understood componentwise and $R_{j,k}=\delta_{j,k+1} + \delta_{j+1,k}$. This has solution 
 \begin{eqnarray}\label{eqn:lrbound}
  C_B(x,t) \leq e^{-8(\gamma-8\|H_0\|)t} \sum_j \left(e^{32\|H_0\|Rt}\right)_{x,j} C_B(j,0).
 \end{eqnarray}
This is the main result of this section: a new Lieb-Robinson bound for our class of models. The key idea here is that the first term can (provided the noise $\gamma$ is large enough) exponentially suppress information propagation, whilst the second term can cause ballistic propagation of information. We now analyse the interplay between these two effects:

{\it Discussion of the new bound ---} Noting that $\|R\|=2$, it is clear that if $\gamma < 16\|H_0\|$ then our bound is exponentially growing and the original Lieb-Robinson bound is superior to ours: it
limits us only to ballistic propagation of information. If, however, $\gamma > 16\|H_0\|$ then the right-hand-side of our new bound (\ref{eqn:lrbound}) is negligible if either (i) $|j-k| \geq
t\kappa_\varepsilon$ for some constant $\kappa_\varepsilon>0$ or (ii) $t \geq t_\varepsilon =\log{({C}/{\varepsilon})} /\Gamma$ where $C=\sum_j C_B(j,0)$ and $\Gamma=8\gamma + 128\|H_0\|$. Equivalently, the right hand side of the new bound is non-negligible only when $|j-k| < \kappa_\varepsilon t_\varepsilon$, which is to say that non-negligible correlations can propagate by at most a constant number of sites.

 \begin{figure}
   \centering
   \includegraphics[bb=0 0 240 240, width=\columnwidth]{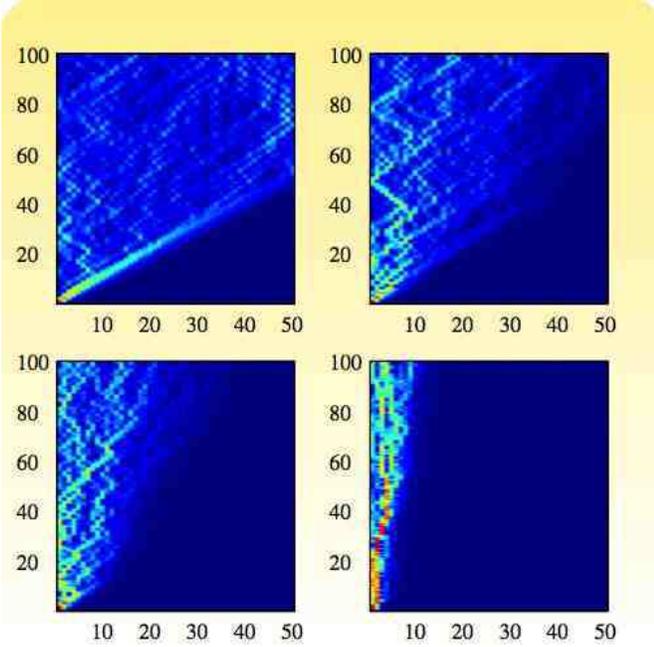}
   \caption{Illustration of the dynamics of the correlation function
   $c_{j,1}(\xi,t) = \langle\Omega| a_{j}(0) a^\dag_1(t)|\Omega\rangle$ generated by Eq.~(\ref{eq:noisexx}) for a 50 site chain for a specific realisation of the fluctuating field $\xi^z_j(t)$. The horizontal axis is site number $j$ and the vertical axis is the time (for $\gamma = 0.05$, $\gamma = 0.1$, $\gamma = 0.2$ and $\gamma = 0.5$).}
   \label{fig:numerics}
 \end{figure}
 
{\it XX spin chain ---} In this section we abandon the Lieb-Robinson commutator in favor of various correlation functions as they are easier to calculate in this specific scenario. In the previous section we studied a noisy field fluctuating in both strength and direction but here we consider a fluctuating field oriented in the  $z$-direction. For convenience we study the XX model, although one can expect similar results to hold for other spin chain models with nearest-neighbor interactions and a fixed field direction. We begin by applying the \emph{Jordan Wigner transformation} to map our system of qubits into a system of spinless fermions 
 \begin{equation}\label{eq:noisexx}
  H(\xi,t) = \sum_j \left(\adagop{j}\aop{j+1} + \adagop{j}\aop{j-1} +
  \xi^z_j(t)\adagop{j}\aop{j}\right)
 \end{equation}
The sum is over an infinite chain or a finite ring (in which case we impose periodic boundary conditions). In order to study the propagation of information through the system we introduce \emph{disorder dependent correlation functions} which are the probability amplitude for a fermion to hop from one site to another in a given time interval
 \begin{equation*}
  c_{j,k}(\xi,t) = \bra{\Omega} \aop{j} U(\xi,t) \adagop{k} \ket{\Omega}
 \end{equation*}
where $\ket{\Omega}$ is the vacuum. We further define \emph{ensemble averaged correlation functions}  $c_{j,k}(t) =\ee_\xi  c_{j,k}(\xi,t) =
\bra{\Omega} \aop{j}(t)\adagop{k}\ket{\Omega}$ where $\aop{j}(t) =\ee_\xi  U^\dag(\xi,t) \aop{j} U(\xi,t)$ is the ensemble averaged annihilation operator in the Heisenberg picture. Using Eq.\ (\ref{eqn:master1}) we form a differential equation for $\text{tr}[\aop{j}(t)\rho]$, yielding $ \partial_t\aop{j}(t) = -i(\aop{j+1}(t) + \aop{j-1}(t)) - \gamma \aop{j}(t)$. Introducing  $\textbf{\emph{a}}(t) = (\aop{1}(t), \aop{2}(t), \ldots)$ allows us to rewrite this as a vector differential equation with solution $\textbf{\emph{a}}(t) = e^{-\gamma t} e^{iRt} \textbf{\emph{a}}(0)$. This in turn implies that the ensemble averaged correlation functions are
 \begin{equation*}
  c_{j,k}(t) = e^{-\gamma t} ( e^{-iRt} )_{j,k}
 \end{equation*}
An analysis similar to that performed previously for the Lieb-Robinson bound reveals that all ensemble averaged correlation functions are exponentially small excepting those for which  $|j-k| < \kappa(\varepsilon,\gamma) = \text{const}$. In other words, the information can---on average---propagate by at most a constant number of sites. In parallel with the results of the first model we have localisation of information but in contrast we have no noise threshold. See Fig.\ \ref{fig:numerics}.

It must be noted that this result holds only when we average over the disorder: specific realisations of the disorder could lead to less localised dynamics. In order to see how far the dynamics of a specific realisation of the noise can stray from the averaged dynamics found above we restrict ourselves to the infinite chain. We introduce the discrete \emph{squared position operator} and the \emph{momentum operator}
 \begin{equation*}
  {x}^2 = \sum_{j=-\infty}^{\infty} j^2 |j\rangle\langle j|,\,
  p=\sum_{j} i\left(
  |j+1\rangle\langle j| - |j\rangle\langle j+1|
  \right)
 \end{equation*}
(where $|j\rangle$ denotes the state vector of a single excitation at site $j$). The Heisenberg picture time derivatives obey equations similar to Eq.\ (\ref{eqn:master1}), e.g.,
$\partial_t {x}^2(t) = i[H_0,{x}^2(t)] - \gamma \mathcal{E}({x}^2(t))$ where  $ \mathcal{E}(M)/2 = M-\sum_j |j\rangle\langle j| M |j\rangle\langle j|$ eliminates the diagonal entries of $M$. When the initial state is a single particle at site $0$, $\rho(0)=|0\rangle\langle 0|$, then
 \begin{equation*}
 	 \langle{x}^2\rangle(t) = \ee_\xi\text{tr}[{x}^2(t)\rho(0)] 
 	 = \frac{2t}{\gamma} + \frac{e^{-2\gamma t}-1}{\gamma^2} =:
 	 f(\gamma,t)
 \end{equation*}
Recalling the trivial bound  $\rho_{j,j}(t) \leq 1$, using translational invariance, and noting that $\langle{x}^2\rangle(t) = \sum_j j^2 \rho_{j,j}(t)$ allows us to conclude that when we place the particle initially at site $k$, $\rho(0)=|k\rangle\langle k|$, then
 \begin{equation}\label{eqn:variance-bound}
  \rho_{j,j}(t) \leq \min \left\{ 1, {f(\gamma,t)}/{|j-k|^2}\right\}
 \end{equation}
One can define \emph{ensemble variance correlation functions} by
 $v_{j,k}(t) =\ee_\xi  |c_{j,k}(\xi,t)|^2 - |\mathbb{E}_\xi  c_{j,k}(\xi,t)|^2$. A little algebra verifies that  $\mathbb{E}_\xi  |c_{j,k}(\xi,t)|^2$ are precisely the diagonal elements $\rho_{j,j}(t)$ when  $\rho(0)=|k\rangle\langle k|$; consequently the $v_{j,k}(t)$ also obey the bound (\ref{eqn:variance-bound}).

This bound implies that it is unlikely that we will stray far from the averaged dynamics given by $c_{j,k}(t)$ for small times $(2t/\gamma)^{1/2}\ll |j-k|$. By employing Chebyshev's inequality which states that for a random variable $X$ with mean $\mu$ and finite variance $\sigma^2$ and for any positive real number $\kappa$ then  $\mathbb{P}\left( \left| X-\mu\right| \geq \kappa\sigma\right) \leq {1}/{\kappa^2}$, we can conclude that there is an effective light cone whose radius grows proportionally to  ${f(\gamma,t)}^{1/2} \sim (2t/\gamma)^{1/2}$. In other words, information can---on average---propagate by a distance proportional to the square-root of the time elapsed, reminiscent of a classical random walk \cite{TailNote}.

To aid understanding of the above result, we express the averaged motion in the Heisenberg picture to find that {\it wave packets stop} in their motion: since $[H_0,p]=0$ and $\mathcal{E}(p)=2p$, the Heisenberg picture momentum obeys $\langle p\rangle (t)=\ee_\xi \text{tr}[p(t)\rho(0)] = e^{-\gamma t} \langle p\rangle (0)$, so an initial excitation  ``localizes'' in momentum space exponentially fast.

{\it Mixing properties of generic fluctuating disorder --- } What is the steady state the dynamics converges to on average for very long times? We now finally link the dynamics under disorder to unitary 
designs
\cite{Design,Harrow}, showing that generically one will obtain the maximally mixed state on average. Consider the following master equation for a spin chain (with $\gamma>0$)
 \begin{equation}\label{eqn:master2}
  \partial_t \rho(t) = -i[\rho(t), H_0] - \gamma \sum_{j}
  [[\rho(t),X_j],X_j]
 \end{equation}
Define the matrix $F$ by $-i [ H_0,B_{\bm\alpha}]= \sum_{\bm\beta} F_{{\bm\alpha},{\bm\beta}} B_{\bm\beta}$, where $B_{\bm\alpha} = \sigma^{\alpha_1}_1 \otimes\cdots\otimes
\sigma^{\alpha_n}_n$ as before. The operators $X_j$ can without loss of generality be taken to be of the form $X_j=x\sigma_j^{0}+y\sigma_j^{z}$, and we write  $I= \{ {\bm\alpha}: \alpha_j\in \{x,y\} \text{ for some } j\}$. If the operators governing the local disorder have full rank and if the submatrix of 
$F$ of entries $F_{{\bm\alpha},{\bm\beta}}$ for which ${\bm\alpha}\not\in I$ and ${\bm\beta}\in I$ is a matrix of maximum rank, then $\lim_{t\rightarrow\infty}\rho(t)=\lim_{t\rightarrow\infty}
\mathbb{E}_\xi  \rho(\xi,t)= (\id/2)^{\otimes n}$ \cite{Proof}, meaning that the state becomes maximally mixed. Almost all Hamiltonians $H_0$ and local fluctuations have this property. Time evolution under fluctuating disorder therefore approximates a unitary $1$-design \cite{Design,Harrow} arbitrarily well for large times and the system ``relaxes'' on average.

{\it Summary and outlook --- } In conclusion we have studied various
regimes of fluctuating on-site disorder (noise) in quantum spin chains and we have identified several regimes of information propagation. These have been compared to the propagation regimes for similar models with both static disorder and no disorder. We have found that in some instances (namely those with low noise levels or fixed field direction) that the localisation due to fluctuating disorder is weaker than that caused by static disorder; conversely we have identified other regimes (those with high levels of unobservable noise) for which 
the localisation is stronger than that caused by static noise. This highlights the complicated nature of disordered quantum systems. Indeed, such localization under time-dependent disorder should also be directly observable in experiments with cold atoms, exploiting similar ideas of fluctuating speckle potentials as used in the seminal experiment of Ref.\ \cite{Aspect}. One could even think of using fluctuating disorder in cold atom systems as a ``disorder filter'', shaping traveling wavepackets in their form or letting them stop.

It is a simple matter to extend the results above by replacing the
intrinsic Hamiltonian $H_0$ with a time dependent one $H_0(t)$: the
same master equation is obeyed. This would allow us to implement quantum logic gates on neighboring qubits hence providing a means
with which to accomplish quantum computation. The important point is
that this provides a more realistic model of noise than has been hitherto used by many authors, including Refs.\ \cite{Buh-et-al06}: the usual scheme is to assume perfect gate operation followed by an instantaneous error; our model allows one to analyse systems where the
qubits are subject to noise and decoherence at all times and in particular whilst a gate is being applied.

{\it Acknowledgements ---} 
This work was supported by the EPSRC, the Nuffield Foundation, the
EU (QAP, COMPAS), and the EURYI.


\begin{thebibliography}{99}
\bibitem{TailNote}
	Note also that Chebyshev's inequality implies that there is
	power law decay of correlations outside this square-root light
	cone which contrasts the exponential decay exhibited by the 
	linear and logarithmic light cones discussed above.
\bibitem{Proof}
	This can be shown by considering the differential equation in the coefficients of the Pauli operator basis \cite{Harrow}, noting when the only fixed point corresponds to the maximally mixed state. One requires the following: let $A$ be real and antisymmetric and $E$ be a positive matrix, $A=\left[\begin{smallmatrix}B & C\\ -C^T & D\end{smallmatrix}\right]$  $E=\left[\begin{smallmatrix}0 & 0\\0 & F\end{smallmatrix}\right]$; if both $F$ and $C$ have maximum rank then $\text{re}(\lambda_j(A-E))<0$ for all $j$.
\end{thebibliography}
\end{document}